\documentclass[iop,tighten,12pt]{emulateapj}

\usepackage{amsmath}
\usepackage{lipsum}

\usepackage{natbib}
\usepackage{aas_macros}
\newcommand{\Cloudy}{\textsc{Cloudy}}

\font\manual=manfnt at 7pt \def\dbend{\hbox{\raise0.9ex\hbox{\manual\char127\hspace{0.6em}}}}

\newcounter{INTERNALionstage}

\def\gtsim{\mathrel{\hbox{\rlap{\hbox{\lower4pt\hbox{$\sim$}}}\hbox{$>$}}}}
\def\lesssim{\mathrel{\hbox{\rlap{\hbox{\lower4pt\hbox{$\sim$}}}\hbox{$<$}}}}

\def\cm{{\rm\thinspace cm}}

\def\micron{\hbox{$\mu$m}}

\def\pcc{{\rm\thinspace cm^{-3}}}

\def\pscm{\mbox{$\cm^{-2}\,$}}

\def\pscm{\mbox{$\cm^{-2}\,$}}

\def\hb{\mbox{{\rm H}$\beta$}}
\def\hi{\mbox{{\rm H~{\sc i}}}}
\def\hii{\mbox{{\rm H~{\sc ii}}}}
\def\hei{\mbox{{\rm He~{\sc i}}}}

\DeclareMathAlphabet{\vib}{OML}{cmm}{m}{it}
\newcommand{\denu}{\ensuremath{n_u}}
\newcommand{\denl}{\ensuremath{n_l}}
\newcommand{\gu}{\ensuremath{g_u}}
\newcommand{\gl}{\ensuremath{g_l}}

\newcommand{\nel}{\ensuremath{n_e}}

\newcommand{\Aul}{\ensuremath{A_{ul}}}
\newcommand{\Bul}{\ensuremath{B_{ul}}}
\newcommand{\Blu}{\ensuremath{B_{lu}}}

\newcommand{\qul}{\ensuremath{q_{ul}}}

\newcommand{\etanu}{\ensuremath{\eta_\nu}}
\newcommand{\nucen}{\ensuremath{\nu_0}}

\newcommand{\jnu}{\ensuremath{j_\nu}}
\newcommand{\kappanu}{\ensuremath{\kappa_\nu}}

\newcommand{\Inu}{\ensuremath{I_\nu}}
\newcommand{\Jnu}{\ensuremath{J_\nu}}
\newcommand{\Iext}{\ensuremath{I^\mathrm{ext}}}
\newcommand{\Jext}{\ensuremath{J^\mathrm{ext}}}

\newcommand{\Inutot}{\ensuremath{I_\nu^\mathrm{tot}}}
\newcommand{\Jnutot}{\ensuremath{J_\nu^\mathrm{tot}}}
\newcommand{\Inuemg}{\ensuremath{I_\nu^\mathrm{emg}}}

\newcommand{\taunu}{\ensuremath{\tau_\nu}}
\newcommand{\pnu}{\ensuremath{p_\nu}}

\newcommand{\pe}{\ensuremath{p_e}}

\newcommand{\rpos}{\ensuremath{\mbox{\boldmath $r$}}}
\newcommand{\n}{\ensuremath{\mbox{\boldmath $\hat{n}$}}}
\newcommand{\ofr}{\ensuremath{(\mbox{\boldmath $r$})}}
\newcommand{\ofn}{\ensuremath{(\n)}}
\newcommand{\ofrn}{\ensuremath{(\mbox{\boldmath $r$, $\hat{n}$})}}
\newcommand{\ofrmn}{\ensuremath{(\mbox{\boldmath $r$, $-\hat{n}$})}}

\newcommand{\chianti}{\textsc{CHIANTI}}
\newcommand{\lamda}{\textsc{LAMDA}}

\begin{document}

\title{Effects of External Radiation Fields on Line Emission --- Application to Star-forming Regions}
\shorttitle{External Radiation \& Line Emission}
\shortauthors{Chatzikos et al.}

\author{
Marios Chatzikos\altaffilmark{1},
G. J. Ferland\altaffilmark{1},
\and
R. J. R. Williams\altaffilmark{2},
\and
P. A. M. van Hoof\altaffilmark{3},
\and
Ryan Porter\altaffilmark{4}
}
\email{mchatzikos@gmail.com}

\altaffiltext{1}{University of Kentucky, Lexington, KY 40506, USA}

\altaffiltext{2}{AWE plc, Aldermaston, Reading RG7 4PR, UK}

\altaffiltext{3}{Royal Observatory of Belgium, Belgium}

\altaffiltext{4}{Department of Physics and Astronomy and Center for Simulational Physics, University of Georgia, USA}

\begin{abstract}
	A variety of astronomical environments contain clouds irradiated by a combination
	of isotropic and beamed radiation fields.  For example, molecular clouds may be
	irradiated by the isotropic cosmic microwave background (CMB),
	as well as by a nearby active galactic nucleus (AGN).
	These radiation fields excite atoms and molecules and produce emission in different ways.
	We revisit the escape probability theorem and derive a novel expression that
	accounts for the presence of external radiation fields.  We show that when the field is isotropic the
	escape probability is reduced relative to that in the absence of external radiation. 
	This is in agreement with previous results
	obtained under ad hoc assumptions or with the two-level system,
	but can be applied to complex many-level models of atoms or molecules.
	This treatment is in the development version of the spectral synthesis code \Cloudy{}.
	We examine the spectrum of a Spitzer cloud embedded in the local
	interstellar radiation field, and show that about 60 percent of
	its emission lines are sensitive to background subtraction.
	We argue that this geometric approach could provide an additional
	tool toward understanding the complex radiation fields of starburst
	galaxies. 
\end{abstract}


%%%%%%%%%%%%%%%%%%%%%%%%%%%%%%%%%%%%%%%%%%%%%%%%%%%%%%%%%%%%%%%%%%%%%%%%%%%%%%%%
\section{Introduction}\label{sec:intro}
%%%%%%%%%%%%%%%%%%%%%%%%%%%%%%%%%%%%%%%%%%%%%%%%%%%%%%%%%%%%%%%%%%%%%%%%%%%%%%%%

	\par
	Emission-line clouds can be powered by a variety of external radiation fields.
	Examples include \hii\ regions powered by nearby stars or the diffuse interstellar medium (ISM)
	powered by the net emission of the surrounding galaxy.
	The angular distribution of the radiation that powers the cloud has little effect on the ionization
	or temperature of a cell of gas, but the radiation transport and net line emission 
	can be quite different, and these affect the observed spectrum.
	
	\par
	We have updated the plasma simulation code \Cloudy{},
	last reviewed by \citet{CloudyReview13}, to account for
	various types of external radiation fields.
	The fundamental distinction is between a ``beamed'' field, such as that
	produced by a single star or nearby AGN,
	versus a nearly isotropic field such as the CMB or a surrounding starburst.
	
	\par
	In sections \S\S\ref{sec:overview}--\ref{sec:cloudy-impl}
	we derive the formalism needed to treat these two cases.
	The observed emission can be quite different when continuum pumping is important.
	In \S\ref{sec:applications} we present an astrophysical application
	on the emission from the diffuse ISM, such as might be observed from
	regions of a starburst galaxy,
	and outline observational tests which can help determine the geometry of
	the radiation field that powers an astrophysical object.
	We conclude in \S\ref{sec:conclusions}.

%%%%%%%%%%%%%%%%%%%%%%%%%%%%%%%%%%%%%%%%%%%%%%%%%%%%%%%%%%%%%%%%%%%%%%%%%%%%%%%%
\section{An overview of escape probabilities}\label{sec:overview}
%%%%%%%%%%%%%%%%%%%%%%%%%%%%%%%%%%%%%%%%%%%%%%%%%%%%%%%%%%%%%%%%%%%%%%%%%%%%%%%%

\begin{figure}
		\begin{centering}
			\includegraphics[scale=0.5]{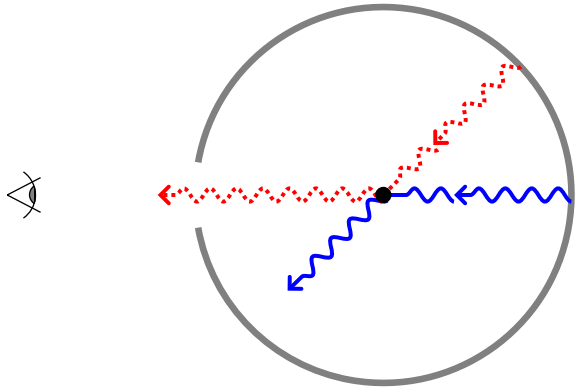}
			\caption{
				An idealized geometry showing the cancellation that occurs with
				an isotropic radiation field.  An atom is taken to lie at the center of
				a surrounding source of radiation.  The observer is to the left.
				The solid (blue) line indicates a photon scattered away from the line
				of sight to the observer.  The dashed (red) line indicates the path of a 
				photon which is scattered into the line of sight, compensating for
				the loss of the original photon.
				\label{fig:geometry}
				}
		\end{centering}
	\end{figure}

	Figure \ref{fig:geometry} shows an idealized geometry 
	of the process we consider.  The observer is to the left and an isolated atom
	is at the center of the cavity.
	If collisional excitation can be neglected, then radiation
	from the surrounding walls is the only source of excitation.
	Assume for the moment that we can neglect stimulated emission produced by the external radiation field.
	Two possible photon paths are indicated.  The solid line indicates the path of a photon
	which is initially directed towards the observer but is scattered away after absorption by the atom.
	This would produce an absorption line were no other photons present.
	The dashed line indicates the path of the photon initially directed transverse to the
	line of sight of the observer, but which is scattered towards the observer after
	encountering the atom.  This would produce an emission line were no other
	photons present.
	In a fully symmetric geometry these two processes completely compensate.
	The atom experiences two excitations but the observer sees no net emission or absorption.
	The same conclusion holds when stimulated emission is taken into account.
	This is in contrast to a beamed continuum, in which an emission
	or absorption line is produced following each photo-excitation.
	
	\par
	Collisions lead to emission or absorption lines
	against the background continuum according to
	Kirchhoff's laws.
	Assuming that the intervening gas is hotter than the background
	continuum, as is the case for the CMB, the intensity of the
	emission line that will form will depend on the gas density.
	Compared to the case of no external radiation,
	the net line emissivity will be reduced.
	The diminution occurs because, although the radiation enhances the
	upper level population and therefore the emissivity, absorption along
	the line of sight is in excess of that gain.
	At the extreme where radiative transitions dominate, the level population
	density ratio should become equal to the Boltzmann ratio at the radiation
	temperature and no emission in excess of the continuum should be possible,
	as discussed above.
	
	\par
	The remainder of this section provides an introduction to
	the escape probability theory and lays the groundwork for
	developments in later sections.
	The escape probability theorem is a commonly
	used approximation in radiative transfer.
	It is invoked to simplify the problem of
	photon propagation through a gaseous medium.
	In principle, photons generated at any point
	in the medium may be absorbed or scattered
	at some different point, or
	they may escape without undergoing further
	interactions.
	These potentially non-local couplings render
	the problem of photon transfer analytically, 
	and in many cases even numerically, intractable.
	
	\par
	The theorem states that the {\em net} number
	of photons created at some position in the
	medium in fact escapes without suffering
	any absorptions, leading to a great simplification
	of the problem.
	Although further assumptions need to be made
	regarding the non-escaping photons in
	order to reach a complete solution to
	a radiative transfer problem,
	the value of the escape probability theorem
	lies in that it decouples the radiative transfer
	and the local statistical state problems.
	
	\par
	The first rigorous proof of the theorem is
	due to \citet{Irons.F78On-the-equality-of-the-mean-escape-probability},
	who considered averages over a unit volume of a gas.
	Later, \citet{Rybicki.G84Escape-probability-methods-pp-21-64} showed that
	the theorem applies to individual rays
	of monochromatic radiation as well,
	that is, the theorem is much more
	detailed than was previously realized. 
	Both derivations take into account
	the radiation field generated by
	the medium itself, the so-called
	{\em diffuse} field, but do not
	consider the effects of external
	radiation.
	
	\par
	Previous work has included external fields in
	two fashions.
	First, in multi-level atomic models
	\citep[e.g.,][]{GoldreichKwan1974}
	a term is introduced in the statistical
	state equations to account for 
	the presence of external isotropic
	radiation.
	Alternatively, calculations that rely on
	the state of a two-level atom embedded in
	isotropic external radiation have employed
	a diminution factor for the emergent intensity.
	In the case of hyperfine structure lines,
	\citet{DSD1998} and \citet{SunyaevDocenko2007}
	have derived a factor of the form
	\begin{equation}
		D(\etanu, \nel, n_c) = \Bigl\{ 1 + (1 + \gu/\gl) (\etanu + \nel/n_c) \Bigr\}^{-1}	\,	,
		\label{eqn:intensity-correction-hfs}
	\end{equation}
	where \etanu{} is the photon occupation number
	at the line frequency, \gu{} and \gl{} are the
	statistical weights of the upper and lower states,
	\nel{} is the electron density, and
	$n_c = \Aul / \qul$ is the critical
	density, with \Aul{} and \qul{} being the Einstein
	coefficient and the downward collisional
	rate coefficient, respectively.
	(Collisions are mainly induced by electron impact
	in these calculations.)
	
	\par
	In this paper we extend previous proofs by
	explicitly accounting for external radiation
	fields, and arrive at a novel expression for
	the escape probability.
	Our expression is similar to that adopted by
	\citet{GoldreichKwan1974}, and leads to diminution
	factors consistent with \citet{DSD1998}
	and \citet{SunyaevDocenko2007}.
	However, unlike the latter, our expression
	may be applied to systems for which the
	critical density may not be defined.
	In fact, in our implementation in the spectral
	code \Cloudy{} \citep[C13; ][]{CloudyReview13},
	all transitions are handled uniformly,
	including those whose level populations
	stem from the solution of a multi-level model.
	
	\par
	In \S\ref{sec:two-level} we present the basic equations of
	the two-level atom, while in \S\ref{sec:basic-esc-prob} we introduce
	some key concepts of the escape probability
	theorem.
	Then, in \S\ref{sec:esc-prob-external} we derive the theorem in the
	presence of external radiation both along
	rays, as well as for volume averages.
	In \S\ref{sec:cloudy-impl} we discuss our implementation
	in \Cloudy.

%%%%%%%%%%%%%%%%%%%%%%%%%%%%%%%%%%%%%%%%%%%%%%%%%%%%%%%%%%%%%%%%%%%%%%%%%%%%%%%%
\section{Two-level Atom}\label{sec:two-level}
%%%%%%%%%%%%%%%%%%%%%%%%%%%%%%%%%%%%%%%%%%%%%%%%%%%%%%%%%%%%%%%%%%%%%%%%%%%%%%%%

	\par
	Although the derivation of the theorem is
	quite general, in the following we will
	adopt the forms of the two-level atom for
	convenience.
	It should be understood, however, that the levels
	referred to below are the upper and lower levels
	associated with any line transition, irrespective
	of its (atomic or molecular) origin.
	
	\par
	The level populations, \denl{} and \denu{}, of a two-level atom 
	in steady-state or statistical equilibrium are determined by
	the interplay between collisional and radiative processes.
	The upward electron collisional excitation rate
	is $C_{lu} = \nel \, q_{lu}$, where \nel{} is the electron density,
	and $q_{lu}$ is the upward collisional rate coefficient, measured in
	cm$^3$~s$^{-1}$, which varies slowly with temperature.
	
	\par
	On the other hand, the induced upward radiative
	excitation rate is given by the $\Blu \, \Jnu$,
	where $\Blu$ is the Einstein coefficient for upward
	transitions, and $\Jnu$ is the radiation field mean
	intensity at frequency $\nu$.
	The induced rates are related to the spontaneous
	deexcitation rate, $\Aul$, by
	\begin{equation}
		\Bul = \frac{\gl}{\gu} \Blu = \frac{c^2}{2 h \nu^3} \Aul	\,	,
		\label{eqn:Einstein-B}
	\end{equation}
	where $\gl$ and $\gu$ are the level statistical weights.
	
	\par
	At this point it is useful to define the photon occupation
	number at $\nu$ as
	\begin{equation}
		\etanu = \frac{c^2 \Jnu}{2 h \nu^3}
			\equiv \Bigl[\exp(h\nu / k_B T_\mathrm{ex}) - 1 \Bigr]^{-1}	\,	,
		\label{eqn:photon-occup}
	\end{equation}
	which also serves as the definition for the
	excitation temperature, $T_\mathrm{ex}$.
	
	\par
	In steady state or statistical equilibrium, the level
	populations at a position in the medium are connected
	by the relation
	\begin{equation}
		\Bigl( C_{lu} + \Blu \Jnu \Bigr) \, \denl\ofr 	=
		\Bigl( C_{ul} + \Bul \Jnu + \Aul \Bigr)	\, \denu\ofr \,	,
		\label{eqn:twolev-standard}
	\end{equation}
	which, in light of the previous equations, may be written as
	\begin{equation}
		\Bigl( C_{lu} + \frac{\gu}{\gl} \etanu \Aul	\Bigr) \, \denl\ofr =
		\Bigl( C_{ul} + (1 + \etanu) \Aul		\Bigr) \, \denu\ofr       \,	.
		\label{eqn:twolev}
	\end{equation}

	\par
	For a transition between the two levels,
	the emissivity, \jnu{},
	the absorption coefficient, \kappanu{},
	and the source function, $S$,
	are given by the relations
	\begin{eqnarray}
		\jnu\ofr	& = &	\frac{h \nu}{4 \pi} \, \Aul \, \denu\ofr \, \psi_{ul}(\nu-\nucen)			\,	, \label{eqn:emiss-coeff}	\\
		\kappanu\ofr	& = &	\frac{h \nu}{4 \pi} \, (\Blu \, \denl\ofr - \Bul \, \denu\ofr) \, \phi_{lu}(\nu-\nucen)	\,	, \label{eqn:absor-coeff}	\\
		S\ofr		& = &	\frac{\jnu\ofr}{\kappanu\ofr} = \frac{\Aul \,\denu\ofr}{\Blu \, \denl\ofr - \Bul \, \denu\ofr}	\,	, \label{eqn:src-func}
	\end{eqnarray}
	respectively.
	The last equation invokes complete redistribution,
	so that the emission and absorption line profiles
	are the same:
	$\psi_{ul}(\nu-\nucen) = \phi_{lu}(\nu-\nucen) = \phi(\nu-\nucen)$.
	This assumption is generally valid in the cores
	of strong resonance lines, as well as in the line
	wings when redistribution due to collisions
	dominates over coherent scattering
	\citep{Jefferies1968}.
	In addition, complete redistribution is appropriate
	for transitions that do not involve the
	ground level, as is the case for subordinate lines
	between excited states.
	
	\par
	For complete redistribution, the source function is
	constant across the line profile, and it may be further
	simplified through equation~(\ref{eqn:Einstein-B}) to
	\begin{equation}
		S = \frac{2h\nu^3}{c^2} \Bigl\{ \frac{\denl}{\denu} \frac{\gu}{\gl} - 1 \Bigr\}^{-1}	\,	.
		\label{eqn:src-func-reduced}
	\end{equation}
	Coupling with equation~(\ref{eqn:photon-occup}),
	the mean intensity may be expressed in terms of the source function
	as
	\begin{equation}
		\Jnu = S \, \etanu \, \Bigl\{ \frac{\denl}{\denu} \frac{\gu}{\gl} - 1 \Bigr\}	\,	.
		\label{eqn:J-interms-S}
	\end{equation}

	\par
	For reference, we note that the first moment in frequency
	of any function, $f(\nu)$, that is symmetric about the line
	center, is found to be
	\begin{align}
		\int_0^\infty d\nu \, \nu \, \phi(\nu-\nucen) \, f(\nu) 
			= {} &	\int_{-\infty}^\infty d\nu \, \nu \, \phi(\nu-\nucen) \, f(\nu)   		\nonumber	\\
			= {} &	\int_{-\infty}^\infty d\nu \, (\nu-\nucen) \, \phi(\nu-\nucen) \, f(\nu)   	\nonumber	\\
			     &	+ \nucen \, \int_{-\infty}^\infty d\nu \, \phi(\nu-\nucen) \, f(\nu)   		\nonumber	\\
			= {} &	\nucen \, \int_0^\infty d\nu \, \phi(\nu-\nucen) \, f(\nu)	\,	,
		\label{eqn:freq-integ-proof}
	\end{align}
	since the first integral in the second step is identically zero.
	Then, the total line emissivity and absorption may be obtained by
	integrating equations~(\ref{eqn:emiss-coeff}) and
	(\ref{eqn:absor-coeff}) over frequency, namely
	\begin{eqnarray}
		j(\mbox{\boldmath $r$})		& = &	\frac{h\nucen}{4\pi} \, \Aul \, \denu\ofr					\label{eqn:emiss-coeff-total}	\\
		\kappa(\mbox{\boldmath $r$})	& = &	\frac{h\nucen}{4\pi} \, \Bigl(\Blu \, \denl\ofr - \Bul \, \denu\ofr \Bigr)	\label{eqn:abs-coeff-total}	\,	.
	\end{eqnarray}

	\par
	The net line emissivity is the number of photons created
	by spontaneous and induced emission, corrected for the
	number of photons absorbed, that is,
	\begin{align}
	        4\pi j^\mathrm{net}\ofr	= {} &	4\pi \, \int d\nu \, (\jnu\ofr - \kappanu\ofr \, \Jnu\ofr)		\nonumber	\\[5pt] 
					= {} &	4\pi \, j\ofr \, 	 						\nonumber	\\
					     &	\times \int d\nu \, \phi(\nu-\nucen) \, \Bigl( 1 - \Jnu\ofr / S\ofr \Bigr) 	\,	,
	        \label{eqn:net-emission}
	\end{align}
	where we have assumed that the radiation field is symmetric
	about the line center, as discussed in \S\ref{sec:esc-prob-external},
	and made use of equations~(\ref{eqn:freq-integ-proof}) and (\ref{eqn:emiss-coeff-total}).
	
	\par
	The final integral of equation~(\ref{eqn:net-emission})
	defines the local {\em net radiative bracket} as
	\begin{equation}
		\rho\ofr \equiv 1 - \frac{1}{S\ofr} \, \int d\nu \, \phi(\nu-\nucen) \, \Jnu\ofr 	\,	,
		\label{eqn:bracket-def}
	\end{equation}
	which may be interpreted as an effective
	reduction of the spontaneous deexcitation
	rate due to the radiation field.
	
	\par
	The bounds of the net radiative bracket are $0 \le \rho \le 1$.
	The lower limit occurs either in pure scattering or
	in local thermodynamic equilibrium, when all processes
	are in detailed balance.
	The upper limit corresponds to the extreme case
	of no diffuse radiation (see next section),
	for which all emitted photons escape.
	However, the net radiative bracket may assume
	negative values due to line-interlocking in a
	multi-level atom \citep{Athay1972}, and values
	greater than unity in masers \citep{Elitzur1992}.

%%%%%%%%%%%%%%%%%%%%%%%%%%%%%%%%%%%%%%%%%%%%%%%%%%%%%%%%%%%%%%%%%%%%%%%%%%%%%%%%
\section{Basic Concepts for the Escape Probability Theorem}\label{sec:basic-esc-prob}
%%%%%%%%%%%%%%%%%%%%%%%%%%%%%%%%%%%%%%%%%%%%%%%%%%%%%%%%%%%%%%%%%%%%%%%%%%%%%%%%

	\par
	A fraction of the line photons that emerge from a volume element
	will be scattered or absorbed as they transverse the medium along
	the line of sight, and will be removed from the beam; the remainder
	will reach the medium boundary in a single flight.
	The fraction of escaping photons typically defines the escape probability,
	which is a function of the optical depth to the unit volume at some
	frequency, and of the geometry of the medium.
	The escape probability along a single ray is denoted by \pnu{},
	and is related to the monochromatic optical depth, $\taunu\ofrn = \tau\ofrn \, \phi(\nu-\nucen)$,
	through the exponential extinction law,
	\begin{equation}
		\pnu\ofrn = \exp\Bigl(-\tau\ofrn \, \phi(\nu-\nucen)\Bigr)			\,	.
		\label{eqn:esc-prob-def}
	\end{equation}
	The average escape probability over the line profile is
	\begin{equation}
		\pe\ofrn = \int_0^\infty	d\nu \, \phi(\nu-\nucen) \, \pnu\ofrn		\,	.
		\label{eqn:esc-prob-freq-avg}
	\end{equation}
	Further averaging over all angles leads to the local mean escape probability,
	sometimes referred to as the local escape factor,
	\begin{equation}
		P_e\ofr = \frac{1}{4\pi} \int d\Omega \, \pe\ofrn				\,	.
		\label{eqn:esc-prob-mean}
	\end{equation}
	
	\par
	The problem of accounting for the lost photons is highly complex
	because it involves interaction between potentially distant regions
	of the emitting medium \citep{Osterbrock1962}.
	The on-the-spot approximation is usually made to bypass that problem,
	which assumes that the non-escaping photons are re-absorbed in situ,
	without further transversing the medium.
	The trapped photons then contribute to the local diffuse field,
	whose mean intensity may be written as
	\begin{equation}
		J^\mathrm{diff}\ofr = \Bigl( 1 - P_e\ofr \Bigr) S\ofr	\,	.
		\label{eqn:local-diffuse-field}
	\end{equation}

	\par
	The main benefit of introducing the concept of escape probability
	is that it decouples the macroscopic state of the medium from
	the local microscopic state, as well as the statistical state
	from the problem of radiative transfer, which can now be solved
	independently of each other.
	
	\par
	The connection to  macroscopic observables may be facilitated by
	considering averages of these processes over a macroscopic
	volume of the medium.
	Indeed, \citet{Irons.F78On-the-equality-of-the-mean-escape-probability}
	showed that emission-weighted averages
	of the net radiative bracket and the escape probability are equal, 
	\begin{equation}
		\langle \rho \rangle_V = \langle P_e \rangle_V	\,	.
		\label{eqn:esc-prob-no-rad}
	\end{equation}
	This result is known as Irons' theorem.
	On the other hand, \citet{Rybicki.G84Escape-probability-methods-pp-21-64} showed
	that this equality applies even to single rays,
	and so it has a more detailed meaning than the energy
	conservation suggested by Iron's treatment.
	We refer to this as Rybicki's theorem,
	namely,
	\begin{equation}
		\Bigl\langle 1 - \frac{\Inu}{S} \Bigr\rangle_\mathrm{ray}
			\equiv	\langle \rho \rangle_\mathrm{ray}
			=	\langle \pnu \rangle_\mathrm{ray}	\,	.
		\label{eqn:monochr-esc-prob-no-rad}
	\end{equation}
	Notice the subtle difference in the definition of the radiative
	bracket between the two treatments.
	In the following, we employ the same symbol for both, and rely on
	the context for disambiguation.
	
	\par
	Note that, in this paper, by emission-weighted integrals we mean integrals
	along a ray or over a volume in which the integrand function at each position
	is multiplied by the emissivity, or by the source function, if the integral
	is over optical depth.
	
	\par
	Below we extend these theorems to include
	the effects of external radiation fields.
	This is important for radio emission lines,
	where correction for the presence of the
	CMB radiation is important.

%%%%%%%%%%%%%%%%%%%%%%%%%%%%%%%%%%%%%%%%%%%%%%%%%%%%%%%%%%%%%%%%%%%%%%%%%%%%%%%%
\section{Escape Probability with External Radiation Fields}\label{sec:esc-prob-external}
%%%%%%%%%%%%%%%%%%%%%%%%%%%%%%%%%%%%%%%%%%%%%%%%%%%%%%%%%%%%%%%%%%%%%%%%%%%%%%%%

	\par
	In the presence of external radiation, the local total radiation field
	along a ray may be expressed as the superposition of the diffuse field,
	and the external field that has penetrated from the remote boundary of
	the medium along the ray, that is,
	\begin{equation}
		\Inutot\ofrn = \Bigl(1 - \pnu\ofrn\Bigr) S\ofr + \pnu\ofrmn \, \Iext\ofn 		\,	.
		\label{eqn:local-rad}
	\end{equation}
	Here, $\pnu\ofrmn$ is the probability
	that a photon will penetrate to the position of interest
	from the remote boundary of the medium along the line of
	sight.
	Scattering of external radiation into the beam is
	implicitly incorporated into the source function,
	$S\ofr$.
	
	\par
	In addition, we assume that the source function
	and the external field do not vary across the line profile.
	Notice, however, that the total intensity is frequency dependent
	regardless of the behavior of these fields, due to the variation
	with frequency of the escape probability (equation~(\ref{eqn:esc-prob-def})).
	
	\par
	Equation~(\ref{eqn:local-rad}) has been derived previously
	in different contexts.
	For instance, \citet{Castor1970} derived a similar equation
	in the study of Wolf-Rayet stars, while \citet{Elitzur1984}
	obtained an analogous result for one-sided illumination in
	a plane-parallel geometry.
	Our treatment is appropriate to any distribution of matter
	embedded in an isotropic radiation field.
	In addition, it can be extended to properly handle anisotropic
	radiation fields, as well as beamed radiation emanating from
	a source, as it will be discussed later.
	
	\par
	Note that equation (\ref{eqn:local-rad}) splits the external radiation 
	along the line of sight from its angular distribution,
	the latter being encapsulated in the source function.
	Averaging over all angles and frequencies, and taking the 
	external field to be isotropic ($\Jext = \Iext\ofn$), one obtains
	\begin{equation}
		\int d\nu \, \phi(\nu-\nucen) \Jnutot\ofr = \Bigl(1 - P_e\ofr\Bigr) S\ofr + P_e\ofr \Jext 		\,	,
		\label{eqn:local-rad-mean}
	\end{equation}
	where we have employed equation~(\ref{eqn:esc-prob-freq-avg}) with the frequency
	average, and equation~(\ref{eqn:esc-prob-mean}) for the probabilities
	along the two opposing directions ($\n$ and $-\n$) with the solid angle average.
	The latter is enabled by the isotropy of the external field,
	and the fact that in equation~(\ref{eqn:esc-prob-mean})
	$\n$ is a dummy variable.
	
	\par
	Substituting into equation~(\ref{eqn:bracket-def}),
	the local net radiative bracket may be written as
	\begin{equation}
		\rho\ofr = P_e\ofr \, \Bigl( 1 - \Jext / S\ofr \Bigr)	\,	.
		\label{eqn:local-rho-Pesc}
	\end{equation}

	\par
	In the following, we adopt Rybicki's treatment and derive the connection
	between averages of the net radiative bracket and the escape probability
	in the presence of external radiation fields, first along a single ray,
	and subsequently over a macroscopic volume.
	An alternative derivation following Irons' treatment may be found in
	Appendix~\ref{app:sec:irons}.
	Note that unless otherwise stated, angle brackets denote emission-weighted
	averages.

	%%%%%%%%%%%%%%%%%%%%%%%%%%%%%%%%%%%%%%%%%%%%%%%%%%%%%%%%%%%%%%%%%%%%%%%%%%%%%%%%
	\subsection{Along Single Rays}\label{sec:Rybicki-proof}
	%%%%%%%%%%%%%%%%%%%%%%%%%%%%%%%%%%%%%%%%%%%%%%%%%%%%%%%%%%%%%%%%%%%%%%%%%%%%%%%%
	
		\par
		Following \citet{Rybicki.G84Escape-probability-methods-pp-21-64},
		we integrate the equation of radiative transfer along a ray directly,
		without the use of the integrating factor, $e^{\taunu}$, to obtain
		the emergent (net) intensity,
		\begin{eqnarray}
			\Inuemg\ofn	& = &	\int dl \, \kappanu\ofr \, (S\ofr - \Inutot\ofrn)			\nonumber	\\
					& = &	\int dl \, \jnu\ofr \, \Bigl(1 - \frac{\Inutot\ofrn}{S\ofr} \Bigr)	\nonumber	\\
					& = &	\langle \rho \rangle_\mathrm{ray} \int dl \, \jnu\ofr			\,	.
			\label{eqn:Iemg-def-rho}
		\end{eqnarray}
		
		\par
		Alternatively, the emergent intensity may be obtained
		by subtracting the incident external radiation along
		the line of sight from the standard solution to the
		radiative transfer problem, which after a few manipulations
		becomes
		\begin{align}
			\Inuemg\ofn	{} = &	\int dl \, \kappanu\ofr \, \pnu\ofrn \, \Bigl(S\ofr - \Iext\ofn\Bigr)		 		\nonumber	\\
					{} = &	\Bigl\langle \pnu\ofrn \, \Bigl(1 - \frac{\Iext\ofn}{S\ofr}\Bigr) \Bigr\rangle_\mathrm{ray}	\nonumber	\\
					     &	\times \int dl \, \jnu\ofr 									\,	.
			\label{eqn:Iemg-Pesc} 
		\end{align}
		
		\par
		Comparison with equation~(\ref{eqn:Iemg-def-rho}) yields
		\begin{equation}
			\langle \rho \rangle_\mathrm{ray}	=	\Bigl\langle \pnu\ofrn \, \Bigl(1 - \frac{\Iext\ofn}{S\ofr}\Bigr) \Bigr\rangle_\mathrm{ray}	\,	,
			\label{eqn:esc-prob-ext-ray}
		\end{equation}
		where the average is taken along a single ray at a single
		frequency.
		This expression generalizes Rybicki's theorem
		to the presence of external radiation fields,
		and reduces to the original formulation in their
		absence.

	%%%%%%%%%%%%%%%%%%%%%%%%%%%%%%%%%%%%%%%%%%%%%%%%%%%%%%%%%%%%%%%%%%%%%%%%%%%%%%%%
	\subsection{Volume Average}\label{sec:vol-avg}
	%%%%%%%%%%%%%%%%%%%%%%%%%%%%%%%%%%%%%%%%%%%%%%%%%%%%%%%%%%%%%%%%%%%%%%%%%%%%%%%%
	
		\par
		We now extend this result over a macroscopic volume.
		We return to equation~(\ref{eqn:Iemg-def-rho}) and notice
		that a differential cross-section, $dA$, along the
		line of sight contributes to the net monochromatic
		emission along a ray an amount
		\begin{eqnarray}
			dA \, dl \, \kappanu\ofr \, (S\ofr - \Inutot\ofrn)	& = &	\nonumber	\\
			dV \, \jnu(\mbox{\boldmath $r$}) \, (1 - \Inutot\ofrn / S\ofr)	\,	.
			\label{eqn:diff-emission}
		\end{eqnarray}

		\par
		It is useful to notice that, according to equation~(\ref{eqn:freq-integ-proof}),
		the integral of the escape probability over frequency may be obtained by using
		$f(\nu) \equiv \pnu\ofrn$, that is,
		\begin{eqnarray}
			\int_0^\infty d\nu \, \nu \, \phi(\nu-\nucen) \, \pnu\ofrn
				& = &	\nucen \, \pe\ofrn	\,	.
			\label{eqn:Lemg-freq-integ-proof}
		\end{eqnarray}
		Then, the emission- and absorption-weighted integrals
		of the frequency-dependent escape probability reduce,
		respectively, to
		\begin{eqnarray}
			\int d\nu \, \jnu\ofr		\, \pnu\ofrn	& = &	j(\mbox{\boldmath $r$}) \, \pe\ofrn		\,	,	\label{eqn:esc-emiss-wgt}	\\
			\int d\nu \, \kappanu\ofr	\, \pnu\ofrn	& = &	\kappa(\mbox{\boldmath $r$}) \, \pe\ofrn	\,	.	\label{eqn:esc-abs-wgt}	
		\end{eqnarray}
		
		\par
		To obtain the total line luminosity over the continuum,
		we integrate over all angles, all frequencies, and over
		the entire volume of the emitting medium
		to obtain
		\begin{eqnarray}
			L^\mathrm{emg}	& = &	\int dV \int d\nu \, \jnu(\mbox{\boldmath $r$}) \, \int d\Omega \, (1 - \Inutot\ofrn/S\ofr)	\nonumber	\\
					& = &	4\pi \, \int dV \int d\nu \, \jnu(\mbox{\boldmath $r$}) \, (1 - \Jnutot\ofr/S\ofr)		\nonumber	\\
					& = &	4\pi \, \bigl\langle \rho \bigr\rangle_V \, \int dV \, j(\mbox{\boldmath $r$})				\,	,
			\label{eqn:Lemg-rho}
		\end{eqnarray}
		where we have used equations~(\ref{eqn:net-emission}) and
		(\ref{eqn:bracket-def}), and the definition
		\begin{equation}
			\bigl\langle \rho \bigr\rangle_V = \frac{\int dV \, j(\mbox{\boldmath $r$}) \, \rho\ofr}{\int dV \, j(\mbox{\boldmath $r$})}	\,	.
			\label{eqn:def-rho-vol-avg}
		\end{equation}
		Notice that equation~(\ref{eqn:Lemg-rho}) is valid regardless
		of the isotropy of the external radiation.
		Also notice that, in the above and in what
		follows below, the solid angle integration
		is done in a reference frame whose origin
		is at position $\rpos$.
		
		\par
		On the other hand, if we substitute $\Inutot\ofrn$ from
		equation~(\ref{eqn:local-rad}), and notice that because
		of the integration over solid angle we can use $\pnu\ofrn$
		instead of $\pnu\ofrmn$, we obtain
		\begin{eqnarray}
			L^\mathrm{emg}	& = &	\int dV \int d\Omega \int d\nu \, \jnu\ofr \, \pnu\ofrn \, (1 - \Iext\ofn/S\ofr)					\nonumber	\\
					& = &	\int dV \int d\Omega \, (1 - \Iext\ofn/S\ofr) \, \int d\nu \, \jnu\ofr \, \pnu\ofrn					\nonumber	\\
					& = &	4\pi \, \Bigl\langle P_e\ofr \, \Bigl(1 - \Jext/S\ofr\Bigr)\Bigr\rangle_V \, \int dV \, j(\mbox{\boldmath $r$})		\,	,
			\label{eqn:Lemg-Pesc}
		\end{eqnarray}
		where we have assumed that the external radiation is isotropic,
		and we have made use of equations~(\ref{eqn:esc-prob-freq-avg}),
		(\ref{eqn:esc-prob-mean}) and (\ref{eqn:esc-emiss-wgt}).
		
		\par
		Comparing equations~(\ref{eqn:Lemg-rho}) and (\ref{eqn:Lemg-Pesc}) we arrive at
		the volume-averaged equivalent of equation~(\ref{eqn:esc-prob-ext-ray}), that is,
		\begin{equation}
			\langle \rho \rangle_V	=	\Bigl\langle P_e\ofr \Bigl(1 - \frac{\Jext}{S\ofr}\Bigr) \Bigr\rangle_V	\,	.
			\label{eqn:esc-prob-ext-ray-vol}
		\end{equation}

		\par
		An alternate derivation that follows closely
		\citet{Irons.F78On-the-equality-of-the-mean-escape-probability}
		is presented in Appendix~\ref{app:sec:irons}.

	%%%%%%%%%%%%%%%%%%%%%%%%%%%%%%%%%%%%%%%%%%%%%%%%%%%%%%%%%%%%%%%%%%%%%%%%%%%%%%%%
	\subsection{Anisotropic Radiation Fields}\label{sec:aniso}
	%%%%%%%%%%%%%%%%%%%%%%%%%%%%%%%%%%%%%%%%%%%%%%%%%%%%%%%%%%%%%%%%%%%%%%%%%%%%%%%%
		
		\par
		Let us now treat the more general case of anisotropic radiation
		fields, in which the intensity varies with direction.
		
		\par
		In the case of single rays, equation~(\ref{eqn:esc-prob-ext-ray})
		still holds.
		For beamed radiation away from the line of sight,
		this reduces to Rybicki's theorem, equation~(\ref{eqn:monochr-esc-prob-no-rad}).
		However, in this case, the external radiation
		will affect the line source function.
		
		\par
		For volume aggregates, equation~(\ref{eqn:Lemg-rho}) still holds,
		while equation~(\ref{eqn:Lemg-Pesc}) may be written as
		\begin{align}
			L^\mathrm{emg} = {} &  	\int dV \int d\Omega \int d\nu \, \kappanu\ofr 							\nonumber	\\[-2pt] 
				            &  	\times \Bigl(S\ofr \, \pnu\ofrn \, - \pnu\ofrmn \, \Iext\ofn\Bigr)				\nonumber	\\[-2pt]
				       = {} &  	\int dV \int d\Omega \int d\nu \, \jnu\ofr \, \pnu\ofrn						\nonumber	\\[-2pt]
				            &  	- \int dV \int d\Omega \, \Iext\ofn \, \int d\nu \, \kappanu\ofr \, \pnu\ofrmn			\nonumber	\\[-2pt]
				       = {} &  	4\pi \int dV \, j\ofr \, P_e(\mbox{\boldmath $r$})						\nonumber	\\[-2pt]
				            &  	- \int dV \, \kappa\ofr \, \int d\Omega \, \pe\ofrmn \, \Iext\ofn				\nonumber	\\[-2pt]
				       = {} &  	4\pi \Bigl\langle P_e\ofr \, \Bigl( 1- \frac{\Jext_\Omega\ofr}{S\ofr} \Bigr) \Bigr\rangle_V
							\int dV \, j\ofr	\,	,
			\label{eqn:Lemg-Pesc-aniso}
		\end{align}
		where we have used equation~(\ref{eqn:esc-abs-wgt}), and
		\begin{equation}
			\Jext_\Omega\ofr = \frac{1}{4\pi \, P_e\ofr} \int d\Omega \, \pe\ofrn \, \Iext\ofn
			\label{eqn:aniso-Iext-avg}
		\end{equation}
		is the average external field intensity at position \rpos.
		If the field is isotropic, it reduces to the field
		intensity, \Jext{}.
		
		\par
		Finally, comparison with equation~(\ref{eqn:Lemg-rho}),
		which applies for any angular distribution of the external
		radiation field, leads to
		\begin{equation}
			\langle \rho \rangle_V	=
				\Bigl\langle P_e\ofr \, \Bigl( 1- \frac{\Jext_\Omega\ofr}{S\ofr} \Bigr) \Bigr\rangle_V
		\end{equation}
		For an isotropic field,	this equation reduces to
		equation~(\ref{eqn:esc-prob-ext-ray-vol}).

%%%%%%%%%%%%%%%%%%%%%%%%%%%%%%%%%%%%%%%%%%%%%%%%%%%%%%%%%%%%%%%%%%%%%%%%%%%%%%%%
\section{Implementation in Cloudy}\label{sec:cloudy-impl}
%%%%%%%%%%%%%%%%%%%%%%%%%%%%%%%%%%%%%%%%%%%%%%%%%%%%%%%%%%%%%%%%%%%%%%%%%%%%%%%%
	\par
	We implement this theorem in \Cloudy\footnote{http://svn.nublado.org/cloudy/trunk}
	at revision r8028, and it will be part of the next major release.
	We evaluate emissivities using equations~(\ref{eqn:net-emission})
	and (\ref{eqn:bracket-def}) and appropriately employing
	equation~(\ref{eqn:esc-prob-ext-ray}) for the net radiative bracket,
	depending on the presence of isotropic external fields.
	That is, \Cloudy{} computes angle-integrated emissivities
	reduced by the correct escape probability
	according to
	\begin{align}
		4 \pi j^\mathrm{net}\ofr	{}= &	h\nucen \, \Aul \, P_e\ofr \, \denu\ofr \, (1-\Jext / S\ofr)				\nonumber	\\
						{}= &	h\nucen \, \Aul \, P_e\ofr \, \Bigl[\denu\ofr						\nonumber	\\
						    &	+ \etanu^\mathrm{iso} \, \Bigl(\denu\ofr - \frac{\gu}{\gl} \, \denl\ofr\Bigr)\Bigr]	\,	.
		\label{eqn:cloudy-impl}
	\end{align}
	Here the emissivity, $j^\mathrm{net}\ofr$,
	is computed over the line profile,
	and the first term
	is the standard escape probability result.
	\Cloudy{} performs isotropic continuum subtraction
	by default, although this feature may be deactivated
	at the discretion of the user.
	Note that this expression is identical to that
	employed by \citet{GoldreichKwan1974}.
	
	\par
	Note that in the case of strict thermal equilibrium,
	this expression is zero identically, as expected.
	In the case of optically thin media, the emissivity
	of a line transition experiences a diminution of
	\begin{widetext}
	\begin{eqnarray}
		D(\etanu, \nel, n_c, \Delta E)
			& = & 	{1 + \etanu^\mathrm{iso} \, \Bigl(1- \exp(+ \Delta E/kT)\Bigr) \over
			1 + \etanu^\mathrm{iso} \, (1 + \gu/\gl)
			+ \, (\nel/n_c) \, \Bigl[1+(\gu/\gl)\exp(-\Delta E/kT)\Bigr]}		\,	,
		\label{eqn:intensity-correction}
	\end{eqnarray}
	\end{widetext}
	where $\Delta E$ is the energy separation of the two levels,
	and T is the gas temperature.
	In the limit where $\Delta E$ goes to zero,
	this equation reduces to that of \citet{SunyaevDocenko2007},
	appropriate for hyperfine structure lines
	(equation~(\ref{eqn:intensity-correction-hfs})).

%%%%%%%%%%%%%%%%%%%%%%%%%%%%%%%%%%%%%%%%%%%%%%%%%%%%%%%%%%%%%%%%%%%%%%%%%%%%%%%%
\section{Astrophysical Application}\label{sec:applications}
%%%%%%%%%%%%%%%%%%%%%%%%%%%%%%%%%%%%%%%%%%%%%%%%%%%%%%%%%%%%%%%%%%%%%%%%%%%%%%%%

	\par
	It is useful to investigate whether these
	theoretical considerations have a meaningful
	bearing on observations.
	Although application of the above theorem
	may require detailed modeling of an astrophysical
	object, it is important to note that the
	essential distinction is between isotropic
	and directional (``beamed'') irradiation.
	In the case of isotropic radiation, emergent line
	intensities need to be corrected for the isotropic
	continuum, for example by multiplying with the factor
	of equation~(\ref{eqn:intensity-correction}).
	However, beamed radiation fields do not contribute
	to the continuum, in general, and no corrections are necessary.
	Many geometries will include both types of radiation field.
	
	\par
	As an illustrative example, we quantify the impact of
	the illumination geometry in the emission lines that
	arise from the cold neutral medium of a
	diffuse \citet{Spitzer1978} cloud.
	The gas density is taken constant throughout
	the cloud, with the total hydrogen density set to $1 \, \pcc$,
	and the elemental abundances set to the average of the abundances
	of the warm and cold phases in the interstellar medium
	\citep{Mathis1977,Cowie1986,Savage1996,Meyer1998,MullmanEtAl_CoAbund_98,SnowElAt_Fabund09}.
	The cloud is illuminated by the nearly-isotropic local
	interstellar radiation field, which we
	model with the \Cloudy{} implementation of
	the spectral energy distribution of
	\citet{Black1987}, extinguished by a column
	density of $10^{22} \, \pscm$.
	The cloud is also exposed to galactic background cosmic rays
	that strongly affect its ion-molecular chemistry.
	We use the emission lines of the \chianti{} v7.1
	\citep{Dere.K97CHIANTI---an-atomic-database-for-emission,Landi2012}
	database for elements more massive than boron,
	and the \lamda{} database \citep{Schoier.F05An-atomic-and-molecular-database-for-analysis}
	for atomic and molecular submillimeter lines.
	A minimal script to reproduce our results
	with \Cloudy{} is given in Appendix~\ref{app:sec:script}.
	
	\par
	Note that a number of spectral energy distributions (SEDs) are part of \Cloudy.
	As outlined in Appendix~\ref{app:sec:fields}, these are, by default, divided into beamed or isotropic
	irradiation.
	The implementation allows the geometry of the illumination to be specified by the user
	and overwrite this default.
	
	\par
	We ran this model twice, with the isotropic continuum
	correction applied and disabled, to
	illustrate the difference between an isotropic
	incident field, such as the local interstellar radiation,
	and illumination by a point source that falls away from
	the sightline.
	Note that this model implicitly assumes that the cloud
	is optically thin at all relevant frequencies.
	
	\par
	The cloud has a gas kinetic temperature of about 500 K, which is produced
	by both the radiation field striking it, and background cosmic rays.
	The incident radiation field is heavily extinguished at hydrogen-ionizing
	energies, due to photoelectric extinction in the ISM, but has
	strong components in the soft-X-ray and UV - visible - IR regions.
	The high-energy SED and cosmic rays contribute to both ionization
	and, to a lesser extent, the heating of the gas.
	This maintains a modest level of ionization, with H$^+$/H$ \sim 10^{-2}$,
	producing \hi{} and \hei{} recombination lines.
	The electron fraction is small, so there is a significant component of
	supra-thermal electrons which produce secondary excitation and ionization
	\citep{Dalgarno1999}.
	This produces a significant contribution to UV permitted lines such as \hi{} Ly$\alpha$.
	The UV component of the SED heats the gas through a combination of
	grain and atomic electron photoejection. 
	This results in thermal collisional excitation of low-lying levels,
	producing emission mainly in the IR due to the low temperature.
	Finally, the UV/optical SED can photo-excite permitted lines which
	have lower levels that are the ground or metastable states.
	Due to these different origins, the correction for isotropic pumping will
	be very different for different lines.
	 
	\par
	Of a total of $\sim$400 emission lines across the spectrum,
	the intensities of about 230 of them are reduced at various
	degrees, with about 160 completely removed, when the background
	correction is performed.
	These lines and their intensity decrements
	are presented in Figure~\ref{fig:decrement}.
	For each line, designated in the bottom panel,
	the top panel shows the intensities for the
	two runs as attached bars, with the left (red)
	showing the beamed case where no correction
	is performed, and the right (green) showing
	the isotropic case.
	The decrement, defined as $\delta = 1 - I_\mathrm{iso}/I_\mathrm{beam}$,
	is shown in the bottom panel.
	A decrement of 0 means there is no difference
	between the two intensities, while a value of
	1 means that there is no emission when the
	correction is applied.
	For guidance the $\hb$ intensity is also indicated.
	
	\par
	Lines are produced by collisional excitation, recombination, and pumping.
	Only lines with a significant pumped contribution will be affected.
	Highly excited recombination lines such as the optical \hi{} and \hei{}
	lines are hardly affected.  
	The \hi{} Ly$\alpha$ line has a significant pumped component, and its
	intensity is lowered by about 50 percent when the correction is performed.
	Optical atomic lines such as Na D are mainly photo excited so their
	intensity falls greatly.
	Various lines are affected in various ways, depending on how they are excited.
	
	\par
	First of all, it is important to note that this effect operates
	across the {\em entire} spectrum.
	Most of the lines are reduced by nearly 100 percent when
	the background subtraction is performed, suggesting that the
	density populations of the levels involved in each line
	transition are dominated by radiative pumping.
	Many of the lines are faint even when no correction is
	applied.
	Then, a useful observational application would require the
	identification of a number of radiatively excited,
	relatively bright emission lines, which would be used to
	constrain the illumination geometry.
	
	\par
	This method could be used in connection
	to starburst and ultraluminous infrared galaxies (ULIRGs).
	The origin of the infrared emission has been
	debated for the last 30 years, with radiation from star formation
	and dust-enshrouded active galactic nuclei (AGNs)
	both being able to account for the observations.
	Diagnostic diagrams that exploit the hardness of
	the radiation field have been devised
	\citep[e.g.,][]{Genzel2000} to separate
	the relative contribution of these mechanisms.
	
	\par
	As a complementary diagnostic, we propose that
	emission in a suitable set of optically thin lines
	could be used to further constrain the importance
	of each mechanism.
	This is founded on the recognition that
	the ambient radiation field in starburst
	galaxies should be largely isotropic for
	any position in the disk of a galaxy,
	while that of an AGN should be beamed.
	Thus, if the majority of the infrared luminosity
	is due to star formation, emission in radiatively
	excited lines will be generally suppressed.
	On the other hand, if these lines arise from fluorescence
	by the AGN, their emission should be enhanced.
	Detailed modeling is required to determine the
	effectiveness and applicability of this geometric
	method.

	%%%%%%%%%%%%%%%%%%%%%%%%%%%%%%%%%%%%%%%%%%%%%%%%%%%%%%%%%%%%%%%%%%%%%%%%%%%%%%%%
	%                                    FIGURE                                    %
	%%%%%%%%%%%%%%%%%%%%%%%%%%%%%%%%%%%%%%%%%%%%%%%%%%%%%%%%%%%%%%%%%%%%%%%%%%%%%%%%
	\begin{figure*}
		\begin{centering}
			\includegraphics[scale=0.65,angle=270]{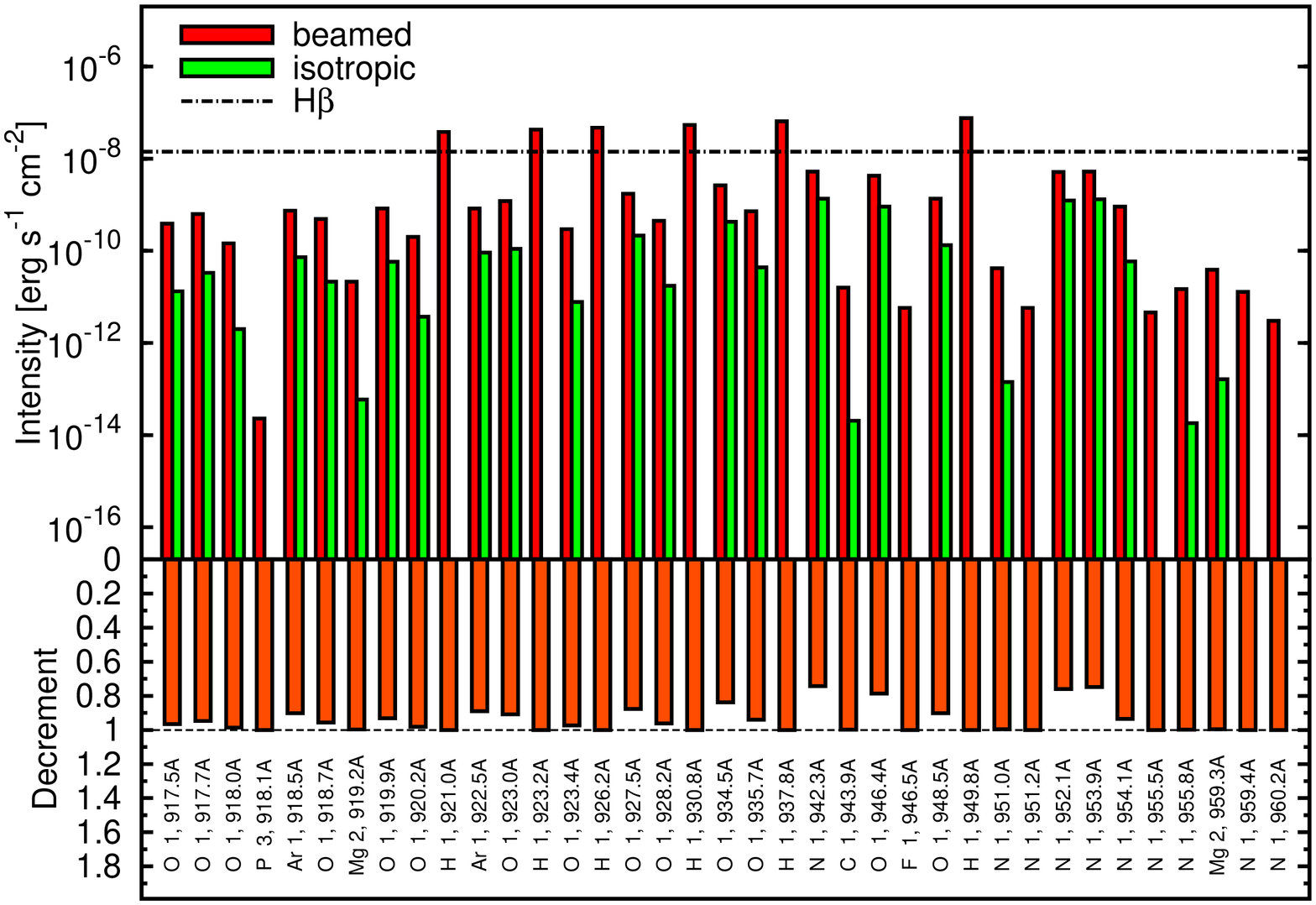}
			\includegraphics[scale=0.65,angle=270]{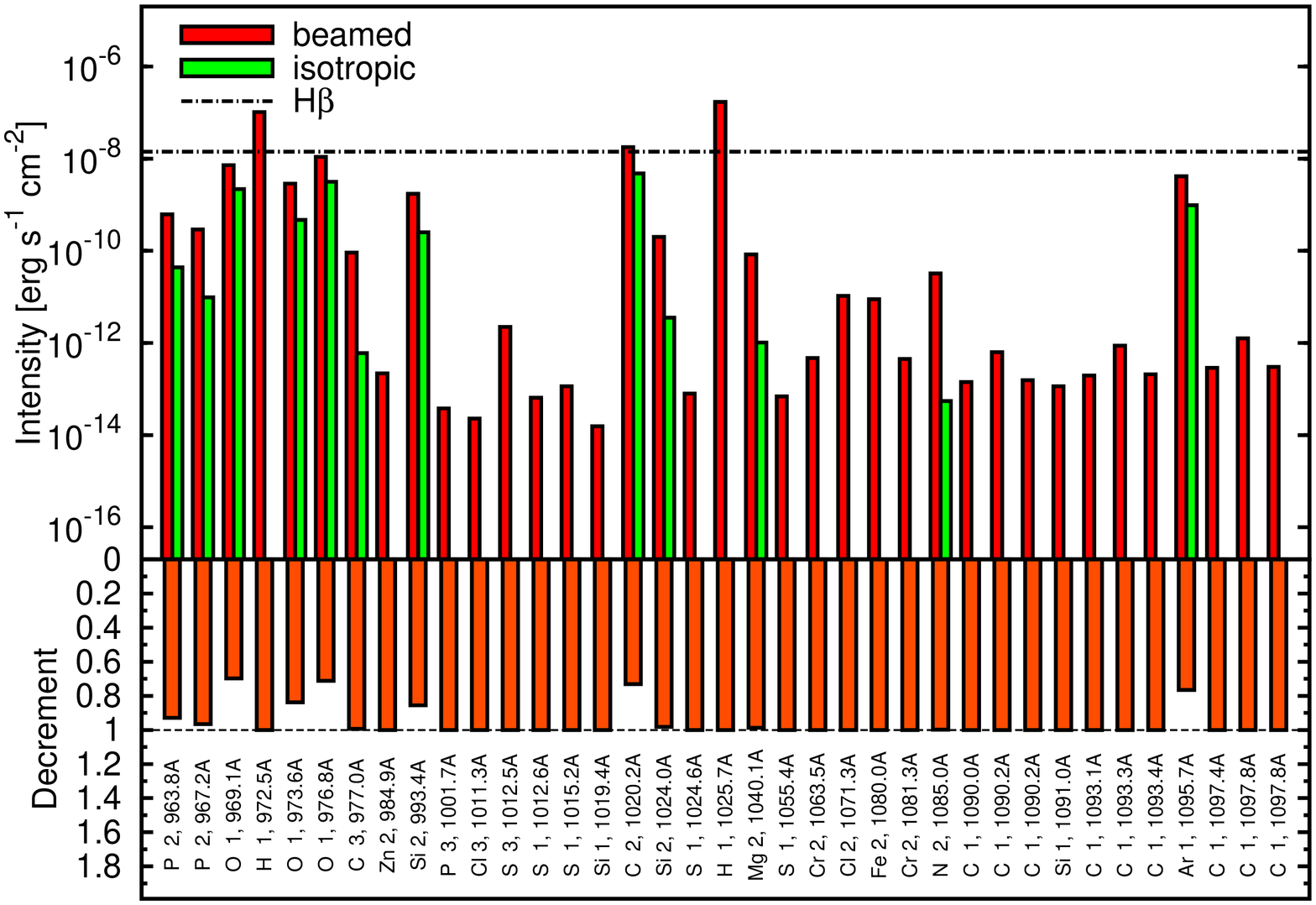}
			\caption{
				Lines affected by the irradiation geometry from the ultraviolet
				to the infrared.
				{\em (Top panel)}
				Intensity obtained by not correcting for the incident continuum,
				``beamed'' case, and by applying the corrections discussed in
				\S\ref{sec:cloudy-impl}, ``isotropic'' case.
				The dot-dashed line indicates the $\hb$ intensity computed for our
				model.
				{\em (Bottom panel)}
				Decremental change, defined as $\delta = 1 - I_\mathrm{iso}/I_\mathrm{beam}$,
				such that if the line intensity goes to zero after the continuum
				correction is applied, the decrement is 1.
				The species and wavelength of the transition
				is shown for each set of bars by the vertical
				labels at the bottom of the panel.
				\label{fig:decrement}
				}
		\end{centering}
	\end{figure*}
	%%%%%%%%%%%%%%%%%%%%%%%%%%%%%%%%%%%%%%%%%%%%%%%%%%%%%%%%%%%%%%%%%%%%%%%%%%%%%%%%
	\begin{figure*}
		\addtocounter{figure}{-1}
		\begin{centering}
			\includegraphics[scale=0.65,angle=270]{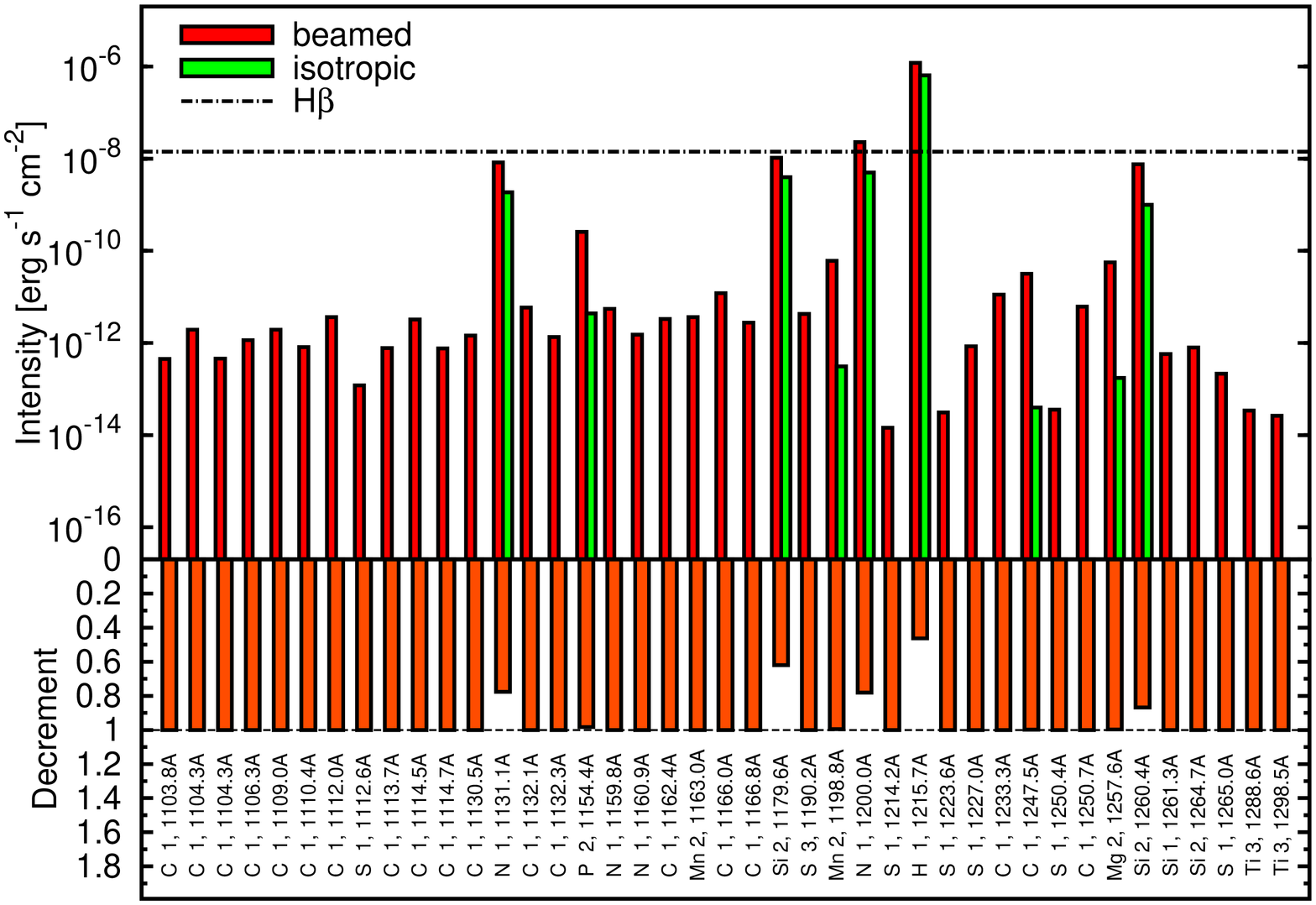}
			\includegraphics[scale=0.65,angle=270]{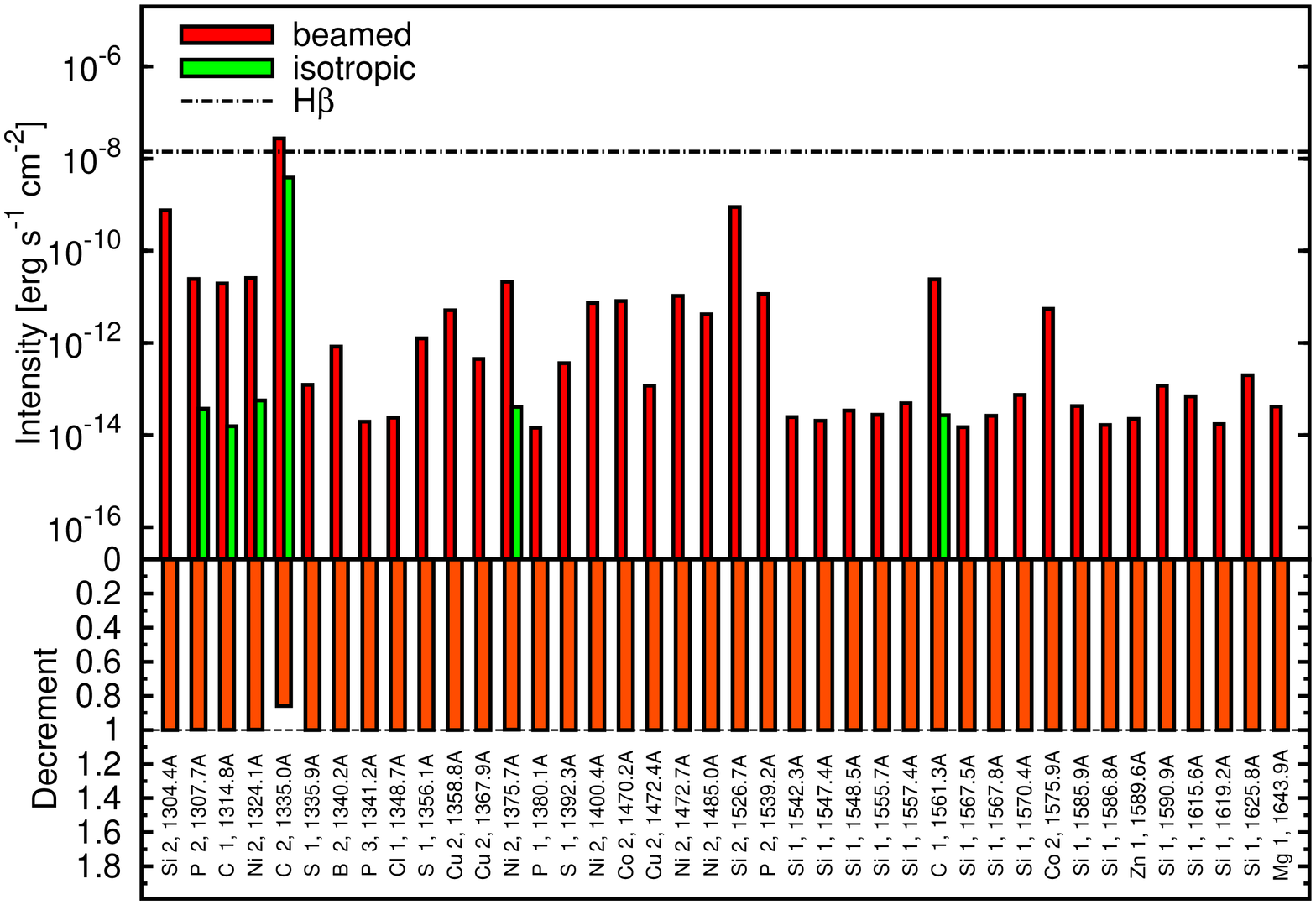}
			\caption{
				{\em (continued)}
				}
		\end{centering}
	\end{figure*}
	%%%%%%%%%%%%%%%%%%%%%%%%%%%%%%%%%%%%%%%%%%%%%%%%%%%%%%%%%%%%%%%%%%%%%%%%%%%%%%%%
	\begin{figure*}
		\addtocounter{figure}{-1}
		\begin{centering}
			\includegraphics[scale=0.65,angle=270]{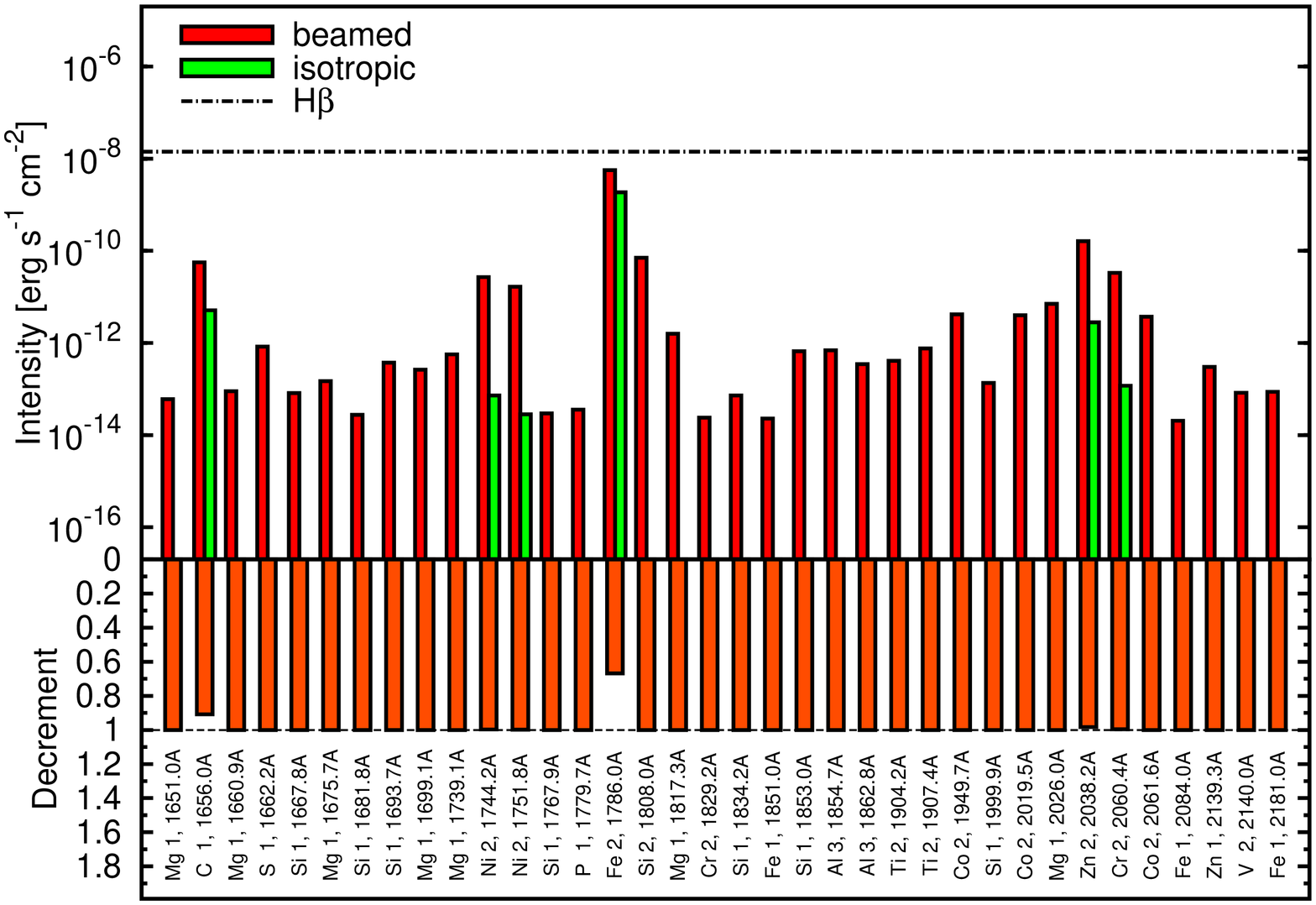}
			\includegraphics[scale=0.65,angle=270]{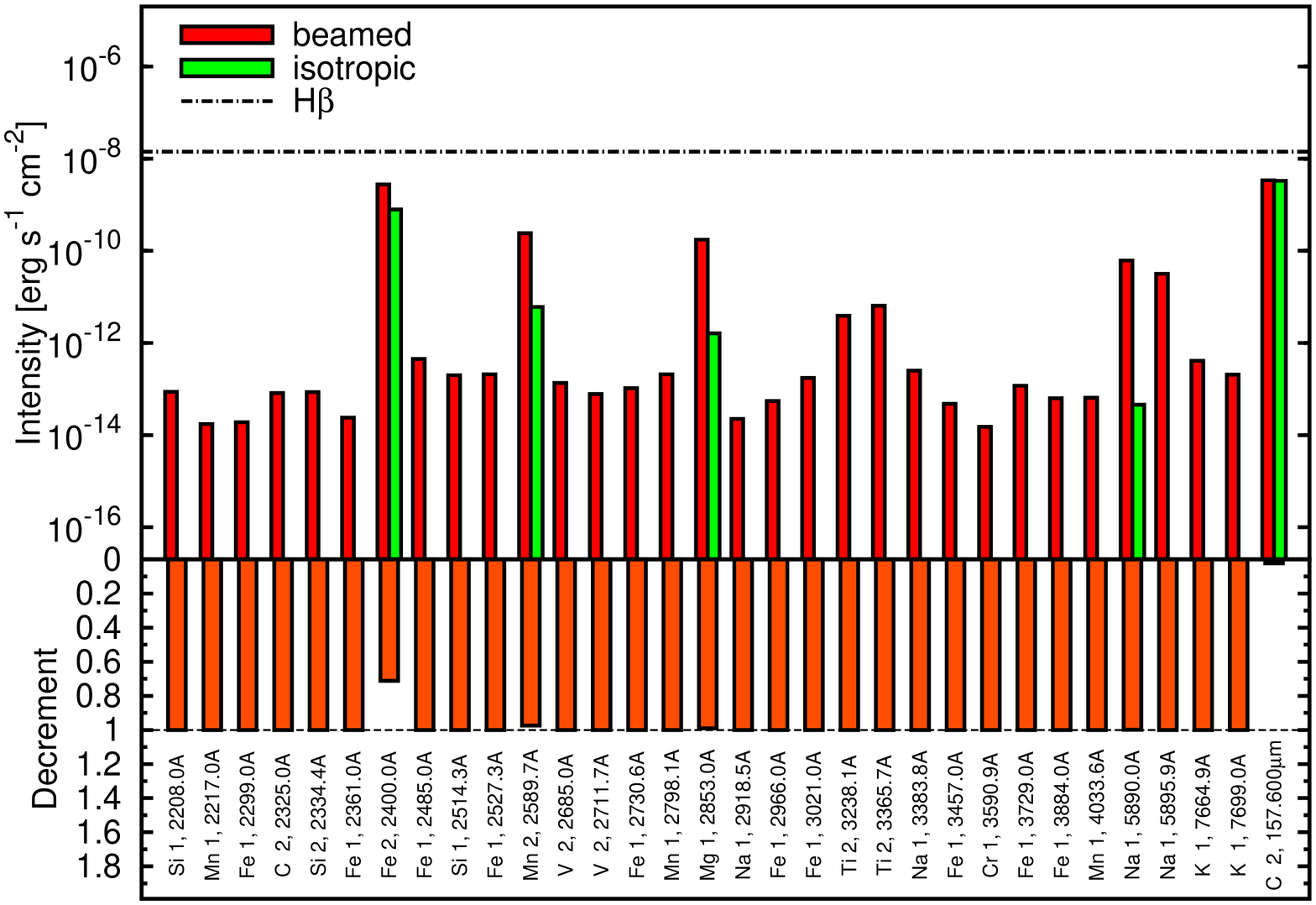}
			\caption{
				{\em (continued)}
				}
		\end{centering}
	\end{figure*}
	%%%%%%%%%%%%%%%%%%%%%%%%%%%%%%%%%%%%%%%%%%%%%%%%%%%%%%%%%%%%%%%%%%%%%%%%%%%%%%%%

%%%%%%%%%%%%%%%%%%%%%%%%%%%%%%%%%%%%%%%%%%%%%%%%%%%%%%%%%%%%%%%%%%%%%%%%%%%%%%%%
\section{Conclusions}\label{sec:conclusions}
%%%%%%%%%%%%%%%%%%%%%%%%%%%%%%%%%%%%%%%%%%%%%%%%%%%%%%%%%%%%%%%%%%%%%%%%%%%%%%%%

	\par
	We have derived the escape probability theorem in the presence
	of external isotropic radiation, and we have shown that this
	effectively leads to a reduction of the net line emissivity
	relative to the unforced case.
	In essence then, the present theorem provides a means
	for a straightforward continuum subtraction for emission lines.
	
	\par
	We have used our implementation of the theorem in \Cloudy{} C13 to study
	the emission lines that arise from a typical Spitzer cloud in the
	presence of an external radiation field.
	We have found that about 60 percent of the emission lines are
	photo-excited and most of them susceptible to external field
	isotropy.
	In light of this finding, we argue that this mechanism may provide
	an additional tool toward quantifying the contributions of the
	isotropic starburst radiation field and the beamed AGN continuum
	in starburst galaxies and ULIRGs.
	A detailed study of the potential offered by this method
	is left for a future paper.

\acknowledgments

	GJF acknowledges support by NSF (1108928; and 1109061), NASA (10-ATP10-0053, 10-ADAP10-0073,
	and NNX12AH73G), and STScI (HST-AR-12125.01, GO-12560, and HST-GO-12309).
	PvH acknowledges support from the Belgian Science Policy office through the ESA PRODEX program.
	\chianti{} is a collaborative project involving George Mason University, the University of Michigan
	(USA) and the University of Cambridge (UK).

%%%%%%%%%%%%%%%%%%%%%%%%%%%%%%%%%%%%%%%%%%%%%%%%%%%%%%%%%%%%%%%%%%%%%%%%%%%%%%%%
\appendix
%%%%%%%%%%%%%%%%%%%%%%%%%%%%%%%%%%%%%%%%%%%%%%%%%%%%%%%%%%%%%%%%%%%%%%%%%%%%%%%%

	%%%%%%%%%%%%%%%%%%%%%%%%%%%%%%%%%%%%%%%%%%%%%%%%%%%%%%%%%%%%%%%%%%%%%%%%%%%%%%%%
	\section{A. Alternate Derivation of Irons' Volume Average}\label{app:sec:irons}
	%%%%%%%%%%%%%%%%%%%%%%%%%%%%%%%%%%%%%%%%%%%%%%%%%%%%%%%%%%%%%%%%%%%%%%%%%%%%%%%%
		\par
		The relation between averages of the net radiative bracket and
		the escape probability for some volume of the medium may also be
		obtained with the energy conservation arguments of \citet{Irons.F78On-the-equality-of-the-mean-escape-probability}.
		
		\par
		Taking the emission-weighted average of equation~(\ref{eqn:local-rho-Pesc})
		over volume, we obtain
		\begin{eqnarray}
			\langle \rho \rangle_V	& = &	\frac{\int dV \, j\ofr \, \rho\ofr}{\int dV \, j\ofr}				\nonumber	\\
						& = &	\frac{\int dV \, j\ofr \, P_e\ofr \, (1 - \Jext / S\ofr)}{\int dV \, j\ofr}	\nonumber	\\
						& = &	\Bigl\langle P_e\ofr \, \Bigl(1 - \frac{\Jext}{S\ofr} \Bigr) \Bigr\rangle_V		\,	;
			\label{eqn:esc-proof}
		\end{eqnarray}
		that is, we recover the same result as with
		Rybicki's ray-based approach.

	\clearpage

	%%%%%%%%%%%%%%%%%%%%%%%%%%%%%%%%%%%%%%%%%%%%%%%%%%%%%%%%%%%%%%%%%%%%%%%%%%%%%%%%
	\section{B. Minimal Script for Astrophysical Application}\label{app:sec:script}
	%%%%%%%%%%%%%%%%%%%%%%%%%%%%%%%%%%%%%%%%%%%%%%%%%%%%%%%%%%%%%%%%%%%%%%%%%%%%%%%%
		\par
		The following script may be used to obtain the results
		presented in Figure~\ref{fig:decrement}.
		
		\begin{verbatim}
			title cloud irradiated by ism background
			#######################
			# Set Continuum
			#######################
			table ism
			illuminate isotropic
			extinguish by a column of 22
			cosmic rays, background
			#######################
			# Set density & abundances
			#######################
			hden 0
			atom chianti "CloudyChianti.ini"
			abundances ism
			#######################
			# Set Geometry
			#######################
			sphere
			stop temperature linear 10
			stop thickness 0.1 linear parsecs
			#######################
			# Other details
			#######################
			iterate to convergence
			#	no lines continuum subtraction
			#######################
			# Output
			#######################
			print last
			print line column
			print line faint -6
		\end{verbatim}

	%%%%%%%%%%%%%%%%%%%%%%%%%%%%%%%%%%%%%%%%%%%%%%%%%%%%%%%%%%%%%%%%%%%%%%%%%%%%%%%%
	\section{C. Spectral Energy Distributions in \Cloudy{} C13}\label{app:sec:fields}
	%%%%%%%%%%%%%%%%%%%%%%%%%%%%%%%%%%%%%%%%%%%%%%%%%%%%%%%%%%%%%%%%%%%%%%%%%%%%%%%%
		\par
		\Cloudy{} C13 includes a number of commands for generating
		radiation fields, as outlined below.
	
		%%%%%%%%%%%%%%%%%%%%%%%%%%%%%%%%%%%%%%%%%%%%%%%%%%%%%%%%%%%%%%%%%%%%%%%%%%%%%%%%
		\subsection{Isotropic Continua}\label{app:sec:isotropic-commands}
		%%%%%%%%%%%%%%%%%%%%%%%%%%%%%%%%%%%%%%%%%%%%%%%%%%%%%%%%%%%%%%%%%%%%%%%%%%%%%%%%
			\par
			\noindent
			\textbf{table SED ``cool.sed''}	\\
			\indent		A cooling flow model based on \citet{Johnstone1992}.
			
			\par
			\noindent
			\textbf{table draine}		\\	
			\indent		The galactic background radiation according to \citet{Draine1996},
					appropriate for simple photon-dominated region (PDR) simulations.
			
			\par
			\noindent
			\textbf{table hm96}		\\	
			\indent		The background continuum radiation at $z=2.16$ of \citet{HaardtMadau96}.
					It does not include the CMB.
			
			\par
			\noindent
			\textbf{table hm05}		\\	
			\indent		The radiation field of Haardt \& Madau (2005, private communication with GJF)
					for any redshift out to $z = 8.9$.  It does not include the CMB.
			
			\par
			\noindent
			\textbf{table ism}		\\	
			\indent		The local interstellar radiation field of \citet{Black1987}.
			
			\par
			\noindent
			\textbf{table stars}		\\
			\indent		External isotropic stellar continua, such as the output of
					Starburst99 \citep{Leitherer1999}.

		%%%%%%%%%%%%%%%%%%%%%%%%%%%%%%%%%%%%%%%%%%%%%%%%%%%%%%%%%%%%%%%%%%%%%%%%%%%%%%%%
		\subsection{Beamed Continua}\label{app:sec:beamed-commands}
		%%%%%%%%%%%%%%%%%%%%%%%%%%%%%%%%%%%%%%%%%%%%%%%%%%%%%%%%%%%%%%%%%%%%%%%%%%%%%%%%
			\par
			\noindent
			\textbf{table SED ``akn120.sed''}\\
			\indent		An AGN continuum according to Peterson (private communication with GJF).
			
			\par
			\noindent
			\textbf{table SED ``crab08.sed''/``crabdavidson.sed''}		\\
			\indent		The continuum of the Crab Nebula, including the pulsar.
					The \textbf{``crab08.sed''} option utilizes the continuum of
					\citet{Atoyan.A96On-the-mechanisms-of-gamma-radiation-in-the-Crab},
					while \textbf{``crabdavidson.sed''} enables the continuum of \citet{Davidson1985}.
					The two differ significantly in the ultraviolet part of the spectrum.
			
			\par
			\noindent
			\textbf{table SED ``rubin.sed''}\\
			\indent		Stellar spectrum that best matches the ionizing continuum of
					the Trapezium stars that illuminate the Orion Nebula.
					This is essentially a \citet{Kurucz1979} stellar spectrum modified by R.H.~Rubin.
			\par
			\noindent
			\textbf{table SED ``xdr.sed''}	\\
			\indent		Continuum of an X-ray dominated region (XDR) according to \citet{Maloney1996}
					in the range 1--100~keV.
			
			\par
			\noindent
			\textbf{table agn}		\\
			\indent		Continuum appropriate for radio-quiet AGN, similar to \citet{Mathews1987},
					except for a break at 10~$\micron$.
					At longer wavelengths the continuum of self-absorbed synchrotron radiation
					is used.
			
			\par
			\noindent
			\textbf{table power law}	\\
			\indent		Power-law continuum appropriate for self-absorbed synchrotron radiation at low
					energies.  At high energies the flux varies as $f_\nu \propto \nu^{-2}$.
					The extent of the mid-range and the spectral index therein are adjustable.
			
			\par
			\noindent
			\textbf{table stars}		\\
			\indent		An extensive number of stellar spectra are available.

%%%%%%%%%%%%%%%%%%%%%%%%%%%%%%%%%%%%%%%%%%%%%%%%%%%%%%%%%%%%%%%%%%%%%%%%%%%%%%%%
%                                 REFERENCES                                   %
%%%%%%%%%%%%%%%%%%%%%%%%%%%%%%%%%%%%%%%%%%%%%%%%%%%%%%%%%%%%%%%%%%%%%%%%%%%%%%%%
\vspace{3cm}
\bibliography{LocalBibliography,bibliography2}

\end{document}